\documentstyle[12pt,epsfig]{article} 
\def\baselinestretch{1.3}
\newcommand{\ba}{\begin{array}}
\newcommand{\ea}{\end{array}}
\newcommand{\bd}{\begin{displaymath}}
\newcommand{\ed}{\end{displaymath}}
\newcommand{\be}{\begin{equation}}
\newcommand{\ee}{\end{equation}}
\newcommand{\bea}{\begin{eqnarray}}
\newcommand{\eea}{\end{eqnarray}}
\newcommand{\sla}[1]{/\!\!\!#1}

\def\a{\alpha}

\def\q2 {q^2}

\def\td {\tilde}
\def\r {\rightarrow}

\def\miss {\hspace{-0.4cm}\slash~~}

\def\rslep {\tilde{e_R}}
\def\rsnu {\tilde{\nu}_R}

\def\snu {\tilde{\nu}}
\def\lslep {\tilde{e_L}}
\def\stau {\tilde{\tau}}
\def\mer {m_{\rslep}}
\def\mmr {m_{\tilde{\mu}_R}}
\def\mml {m_{\tilde{\mu}_L}}
\def\mel {m_{\lslep}}

\def\bt{\begin{table}}
\def\et{\end{table}}

\catcode`@=11 
\def \gsim{\mathrel{\mathpalette\@versim>}}
\def \lsim{\mathrel{\mathpalette\@versim<}}
\def \@versim#1#2{\lower0.4ex\vbox{\baselineskip\z@skip\lineskip\z@skip
     \lineskiplimit\z@\ialign{$\m@th#1\hfil##\hfil$%
     \crcr#2\crcr\sim\crcr}}}
\catcode`@=12 

\begin{document}

\begin{flushright}
{RECAPP-HRI-2010-005}
\end{flushright}

\begin{center}

{\large \textbf {Reconstruction of the left-chiral tau-sneutrino in supersymmetry with a right-sneutrino 
as the lightest supersymmetric particle}}\\[15mm]

Sanjoy Biswas$^{\dagger}$\\
{\em Regional Centre for Accelerator-based Particle Physics \\
     Harish-Chandra Research Institute\\
Chhatnag Road, Jhunsi, Allahabad - 211 019, India}\\[20mm] 

\end{center}

\begin{abstract} 

We have considered a supersymmetric scenario in which the minimal supersymmetric standard model is augmented with 
a right-chiral neutrino superfield for each generation. Such a scenario can have a lightest supersymmetric particle 
(LSP) dominated by the right-chiral sneutrino state and the lighter stau as the next-to-lightest supersymmetric particle 
(NLSP). This can easily be motivated by assuming a high scale framework of supersymmetry breaking like minimal 
supergravity 
(mSUGRA). Due to the extremely small neutrino Yukawa coupling, the decay of the NLSP to the LSP is suppressed and 
consequently the NLSP, here the lighter stau mass eigenstate, becomes stable at the length scale of the detector. 
The collider signal in this case consists of charged tracks of massive stable particles in the muon  chamber. 
Following up on our earlier studies on neutralino and chargino reconstruction in such a scenario, we 
have shown the kinematical information obtained from these charged tracks allows us to reconstruct 
the left-chiral tau-sneutrino as well over a significant region of the mSUGRA parameter space. Two methods for
reconstruction are suggested and their relative merits are discussed.

\end{abstract}

\vskip 3 true cm

{\small \footnoterule{$^{\dagger}$ e-mail: sbiswas@hri.res.in}}

\newpage
\setcounter{footnote}{0}

\def\baselinestretch{1.5}
\section{Introduction}

Physics beyond the Standard Model (SM), often comes with a proliferation of  new particles,
and supersymmetry (SUSY) \cite{Book,Sally} is not an exception to this. Therefore, the determination of 
masses and various other 
properties of these new particles in high-energy collider experiments is of paramount importance. It enables one
not only to discriminate among the multitude of existing models, but also to unveil the fundamental parameters 
in any given scenario. As the Large Hadron Collider (LHC) has already started running, we are closer than ever 
to the fulfilment of this objective.

However, in most SUSY models with conserved 
$R$-parity (defined as $R=(-)^{3B+L+2S}$) the lightest supersymmetric particle (LSP), being neutral and weakly 
interacting, goes without being recorded 
at the detector. This gives rise to large amount of missing transverse energy ($\sla E_T$) which makes the 
reconstruction of the masses of new particles quite difficult.

Although $\sla E_T$ is a canonical signature of most of the supersymmetric theories, one should remember
that the above possibility is not unique. One can have a charged particle as the next-to-lightest supersymmetric 
particle (NLSP) and the decay of it into the LSP may be suppressed, so that the NLSP becomes
stable on the scale of the detector, leaving the charged track of a massive particle in the muon chamber in a collider 
event \cite{Fairbairn:2006gg,Hamaguchi:2006vu,Chou:1999zb,Choudhury:2008gb,sakurai}. From the kinematic information of 
these charged tracks one is able to reconstruct the masses of the superparticles in these scenarios. 
 
This paper is one in a series where we have considered a scenario in which a right-chiral sneutrino is the 
LSP and the NLSP is the superpartner of tau. In the minimal supersymmetric standard model (MSSM) \cite{S.P.Martin1,
Djouadi:1998di}
one can achieve this by the mere addition of a right-handed neutrino superfield for each generation in the MSSM 
spectrum, assuming the neutrinos to be of Dirac type. The superpotential of the MSSM in this case is given 
(suppressing family indices) by 

\be W_{MSSM} = y_l
\hat{L}\hat{H_d}\hat{E^c} + y_d \hat{Q}\hat{H_d}\hat{D^c}+y_u \hat{Q}
\hat{H_u} \hat{U^c}+\mu\hat{H_d}\hat{H_u}+y_\nu \hat{L}\hat{H}_u\hat{\nu}^c_R
\ee 

where $\hat{H_d}$ and $\hat{H_u}$ respectively are the Higgs superfields that give mass
respectively to the $T_{3}=-1/2$ and $T_{3}=+1/2$ fermions. $y's$
are the strengths of Yukawa interactions. $\hat{L}$ and $\hat{Q}$ are
the left-handed lepton and quark superfields respectively, whereas
$\hat{E^c}$, $\hat{D^c}$ and $\hat{U^c}$, in that order, are the right
handed gauge singlet charged lepton, down-type and up-type quark
superfields. $\mu$ is the Higgsino mass parameter.

Such an LSP interacts only through neutrino Yukawa coupling ($y_{\nu}$)
which is $\sim 10^{-13}$ to account for the tiny neutrino masses. The decays of all other sparticles including 
the NLSP into the LSP, is controlled by this Yukawa coupling. Thus the stau becomes long-lived and appears to be 
stable at the detector, drastically changing the signal of SUSY.


It is possible to accommodate such a scenario in a high-scale frame work of SUSY breaking, like minimal 
supergravity (mSUGRA) \cite{mSUGRA} where all the scalar and gaugino masses at low energy are driven by the 
renormalisation group evolution (RGE) 
of the universal scalar ($m_0$) and gaugino mass ($M_{1/2}$) parameters from high scale respectively. All one has to do in 
this framework is to specify  at high scale the values of ($m_0, M_{1/2}, A_0$, $sign(\mu)~{\rm and}~\tan\beta ~=~ 
\langle H_u \rangle/  \langle H_d \rangle$) where, $A_0$ is universal trilinear scalar coupling, $\mu$ is the Higgsino 
mass parameter and $\tan\beta$ is the ratio of the vacuum expectation values of the two Higgs doublets that give masses 
to the up-and down-type quarks. The masses of the right-chiral sneutrinos are also assumed to be evolved from the same 
universal 
scalar mass parameter $m_0$. At one-loop level, the RGE of the right-chiral sneutrino mass parameter is given by
\cite{arkani,S.P.Martin2}
\bea
\frac{dM^2_{\rsnu}}{dt} = \frac{2}{16\pi^2}y^2_\nu~A^2_\nu 
\eea
where $A_{\nu}$ is obtained by the running of the trilinear soft SUSY breaking term $A_0$ and is responsible for 
left-right mixing 
in the sneutrino mass matrix. Due to the small value of $y_{\nu}$, the value of $M_{\rsnu}$ essentially remains anchored 
at $m_0$, whereas 
the other sfermion masses are enhanced at the electroweak scale. Thus, for an ample range of values of the gaugino masses, 
one naturally 
ends up with a sneutrino LSP ($\snu_{1}$) which is substantially dominated by the right-chiral state, because of the fact 
that the mixing 
angle is also controlled by the neutrino Yukawa couplings:

\bea
\tilde{\nu}_1 = - \tilde{\nu}_L \sin\theta + \tilde{\nu}_R \cos\theta
\eea
\noindent
where the mixing angle $\theta$ is given by,
\bea
\tan 2\theta = \frac{2 y_\nu v\sin\beta |\mu\cot\beta -
A_\nu|}{m^2_{\tilde{\nu}_L}-m^2_{\rsnu}}
\eea

It is the NLSP
which determines the nature of collider events. The third generation charged slepton 
often appears as the NLSP in this scenario, due to its larger Yukawa coupling. Since the decay width of the NLSP
is proportional to $y^2_{\nu}$, all supersymmetric cascades in collider events culminate in the pair production 
of the NLSP which decays outside the detector, leaving behind two charged tracks of massive particles in the muon chamber.

The present work is a follow-up of our earlier work on neutralino and chargino reconstruction in supersymmetry with long 
lived stau scenario \cite{Biswas:2009zp,Biswas:2009rba}.
In this work we have concentrated on the mass reconstruction of the heavier tau-sneutrino whose dominant constituent
is the left-chiral state. It is produced in cascade decay of squarks and/or
gluinos via chargino and heavier neutralino decay. The $\snu_{\tau_L}$ thus produced has a substantial branching fraction 
of decaying into a $W\stau$
-pair. Hence, reconstructing the four-momenta of the $W$'s in its hadronic decay mode, it is possible to reconstruct
the heavier mass eigenstate of the tau-sneutrino. Though one can have a $W$ in the final state in association with two 
charged tracks in this long-lived stau scenario, our emphasis nonetheless is on the 
fact that the $W$, 
paired 
with a particular track giving an invariant mass peak offers a definite signature of $\snu_{\tau_L}$ production in the 
SUSY cascade.

This paper is assembled as follows: in the following section we have tried to motivate the choice of benchmark points in 
the context of a mSUGRA framework. In section 3 we have discussed the signal under study and its possible backgrounds. The 
event selection criteria and reconstruction strategy for determining the mass of the heavier $\tau$-sneutrino have also 
been discussed there. The numerical results which comprise a scan over the region of $m_0$-$M_{1/2}$ plane are presented 
in section 4. We conclude in section 5.


\section{The choice of benchmark points}

\begin{table}[htb]
\begin{tabular}{||c||c|c|c|c|c|c||}
\hline
\hline
       & {\bf BP-1}&{\bf BP-2}&{\bf BP-3}&{\bf BP-4}&{\bf BP-5}&{\bf BP-6} \\
\hline
 mSUGRA     &$m_0=100$&$m_0=100$&$m_0=100$&$m_0=100$&$m_0=100$
            &$m_0=100$ \\
 input      &$m_{1/2}=600$&$m_{1/2}=500$&$m_{1/2}=400$&$m_{1/2}=350$
            &$m_{1/2}=325$&$m_{1/2}=325$ \\
            &$\tan\beta=30$ &$\tan\beta=30$&$\tan\beta=30$&$\tan\beta=30$&$\tan\beta=30$
&$\tan\beta=25$ \\
\hline
$\mel,\mml$   &418&355&292&262&247&247\\
$\mer,\mmr$   &246&214&183&169&162&162\\
$m_{\snu_{e_L}},m_{\snu_{\mu_L}}$&408&343&279&247&232&232\\
$m_{\snu_{\tau_L}}$&395&333&270&239&224&226\\
$m_{\snu_{i_R}}$&100&100&100&100&100&100\\
$m_{\stau_1}$&189&158&127&112&106&124\\
$m_{\stau_2}$&419&359&301&273&259&255\\
\hline
$m_{\chi^0_1}$&248&204&161&140&129&129\\
$m_{\chi^0_2}$&469&386&303&261&241&240\\
$m_{\chi^{\pm}_1}$&470&387&303&262&241&241\\
$m_{\tilde{g}}$&1362&1151&937&829&774&774\\
$m_{\tilde{t}_1}$&969&816&772&582&634&543\\
$m_{\tilde{t}_2}$&1179&1008&818&750&683&709\\
$m_{h^0}$ &115&114&112&111&111&111\\
\hline
\hline
\end{tabular}\\
\caption {\small \it Proposed benchmark points (BP) for the 
study of the stau-NLSP scenario
in SUGRA with right-sneutrino LSP. The values of $m_0$ and
$M_{1/2}$ are given in GeV. We have also set $A_0=100~GeV$ and
$sgn(\mu)=+$ for benchmark points under study.}
\label{tab:1}       
\end{table}

In Table 1, we present the benchmark points used in our earlier
studies as well as in this work. 
The mass spectrum is obtained using the spectrum generator 
{\bf \begin{footnotesize} ISAJET 7.78\end{footnotesize}} \cite{isajet}. We have identified those regions of 
$m_0$-$M_{1/2}$ 
plane, where one normally has a $\stau$ LSP, in the absence of a right-chiral sneutrino superfield in an mSUGRA 
scenario. It should be noted, however, that the reconstruction technique we have adopted
is not limited to such a scenario. It can be advocated in all those cases where a stau is the NLSP and its decay
length is large compared to the detector scale \cite{gravitinoLSP,gmsbnlsp,coanni}. The dominantly  right-chiral 
sneutrino state in this 
scenario turns out be a possible dark matter (DM) candidate \cite{Asaka:2005cn}, as it can evade direct DM searches due 
to its minuscule 
Yukawa coupling. The benchmark points we have chosen are not only compatible with the WMAP data \cite{wmap}, but also 
consistent
with other experimental constraints such as $b \rightarrow s\gamma$, correction to the $\rho$-parameter and muon 
($g - 2$) \cite{Amsler:2008zzb,constraints}. The different points in the parameter space are similar in mass 
ordering and thus lead to qualitatively similar
collider signals. However, they correspond to different mass splittings between NLSP and other particles, giving rise
to different kinematics of the final state, which in turn control the reconstructability of various particles. For all 
our benchmark points, the mass hierarchy $m_{\stau}<m_{\chi^0_1}<m_{\snu_{\tau_L}}<m_{\chi^0_2}\approx m_{\chi^{\pm}_1}$ is 
satisfied, which affirms the production of $\snu_{\tau_L}$ in cascade decays of squarks and/or gluinos produced in the 
initial hard scattering.

\section{Signal and backgrounds}

 In order to illustrate that it is possible to reconstruct the left-chiral 
$\tau$-sneutrino in this scenario we have considered the following final state:

\begin{itemize}
\item $\tau_j+W+2\stau+E_{T}\miss+X$
\end{itemize}

\noindent{
where, $\tau_j$ represents a tau jet, the missing transverse energy is denoted by $E_{T}\miss$, 
$W$ symbolizes a $W$-boson that has been identified in its hadronic decay mode\footnote {We have not considered the 
leptonic decay of $W$ since it is difficult to reconstruct the four momenta of the $W$ in its leptonic decay mode due 
to the presence of the invisible neutrino .} 
and all other jets coming from cascade decays are included in X. The collider simulation has been done with a centre of mass energy $E_{cm}$=14 TeV, at two different integrated luminosities of $30 fb^{-1}$ and $100 fb^{-1}$ using the event 
generator {\bf \begin{footnotesize} PYTHIA 6.4.16\end{footnotesize}} \cite{PYTHIA}. A simulation for
the early LHC run at $E_{cm}$=10 TeV and integrated luminosity of $3 fb^{-1}$ has also been predicted. We have used
the parton distribution function CTEQ5L \cite{Lai:1999wy} with the factorisation ($\mu_R$) and renormalisation ($\mu_F$)
scale set at $\mu_R =\mu_F =$average mass of the final state particles. Following are the numerical values of various 
parameters, used in our calculation \cite{Amsler:2008zzb}:}
\\

$~~~~~~~~$ $M_Z=91.187$ GeV,   $M_W=80.398$ GeV,   $M_t=171.4$ GeV  \\
$~~~~~~~~~~~~~$ $M_H=120$ GeV, $\a^{-1}_{em}(M_Z)=127.9$, $~~$  $\a_{s}(M_Z)=0.118$\\


\subsection{Basic idea}

The $\snu_{\tau_L}$ is produced in SUSY cascade predominantly via the decay of lightest chargino ($\chi^{\pm}_1 \r 
\snu_{\tau_L}\tau$) or 
second lightest neutralino ($\chi^{0}_2 \r \snu_{\tau_L}{\bar\nu_{\tau}}$). The corresponding decay branching fraction 
is given in Table-2. 
The momentum information of the charged track enables us to reconstruct the $\snu_{\tau_L}$ mass in this scenario. 
This closely follows our 
earlier studies on neutralino and chargino reconstruction \cite{Biswas:2009zp,Biswas:2009rba}. A procedure for
reconstructing the mass of a left-chiral tau-sneutrino, making occasional use of the earlier results, is outlined here.

\renewcommand{\baselinestretch}{1.1}\selectfont
\begin{table}[htb]
\begin{center}
\footnotesize
\begin{tabular}{||c|c|c|c|c|c||}
\hline
\hline
\multicolumn{2}{||c|}{\bf BP-1}&\multicolumn{2}{c|}{\bf BP-2}&\multicolumn{2}{c||}{\bf BP-3}\\
\hline
$\chi^{0}_2 \r \snu_{\tau_L}{\bar\nu_{\tau}}$(+cc) & $\chi^{\pm}_1 \r \snu_{\tau_L}\tau^{\pm}$ & 
$\chi^{0}_2 \r \snu_{\tau_L}{\bar\nu_{\tau}}$(+cc) & $\chi^{\pm}_1 \r \snu_{\tau_L}\tau^{\pm}$ &
$\chi^{0}_2 \r \snu_{\tau_L}{\bar\nu_{\tau}}$(+cc) & $\chi^{\pm}_1 \r \snu_{\tau_L}\tau^{\pm}$ \\
\hline
 23\% & 24.7\%  & 25\%  & 27.6\%  & 26.6\%  & 32.5\% \\
\hline
\multicolumn{2}{||c|}{\bf BP-4}&\multicolumn{2}{c|}{\bf BP-5}&\multicolumn{2}{c||}{\bf BP-6} \\
\hline
$\chi^{0}_2 \r \snu_{\tau_L}{\bar\nu_{\tau}}$(+cc) & $\chi^{\pm}_1 \r \snu_{\tau_L}\tau^{\pm}$ & 
$\chi^{0}_2 \r \snu_{\tau_L}{\bar\nu_{\tau}}$(+cc) & $\chi^{\pm}_1 \r \snu_{\tau_L}\tau^{\pm}$ & 
$\chi^{0}_2 \r \snu_{\tau_L}{\bar\nu_{\tau}}$(+cc) & $\chi^{\pm}_1 \r \snu_{\tau_L}\tau^{\pm}$ \\
\hline
 22\% & 31.7\%  & 17.8\% & 28.2\% & 17.8\%  & 26.4\%  \\
\hline
\hline
\end{tabular}\\
\caption {\small \it {The branching fractions for $\chi^{0}_2 \r \snu_{\tau_L}{\bar\nu_{\tau}}$(+cc)
 and $\chi^{\pm}_1 \r \snu_{\tau_L}\tau^{\pm}$ for respective benchmark points.}}
\label{tab:1}       
\end{center}
\end{table}
\vspace{-0.5 cm}
The $\snu_{\tau_L}$ has a sizeable decay branching fraction into a $W^{\pm}\stau^{\mp}_1$-pair 
(ranging from $\approx 34\%$ to $84\%$). 
Since the $W$ will always be produced in association with two staus, it is crucial to identify the correct 
$W^{\pm}\stau^{\mp}_1$
-pair and thus avoid the combinatorial background. Since charge identification of a $W$ in 
its hadronic decay mode is difficult, we have adopted the following methods for finding the correct pair:

\begin{enumerate}

\item \underline{Using opposite sign charged tracks (OSCT):} To determine the correct 
$W^{\pm}\stau^{\mp}_1$-pair we have make use of two opposite sign charged tracks. For the signal, one has
$\tau_j+W+2\stau+E_{T}\miss+X$ with one tau in the final state, where the $\stau_1^{\pm}\tau^{\mp}$ pair has 
originated in the decay of a 
neutralino $\chi^0_{1}$ (or $\chi^0_{2}$) and  $\snu_{\tau_L}$ has decayed into a $W^{\pm}\stau^{\mp}_1$ pair. 
The stau-track produced in a $\snu_{\tau_L}$ decay always has the same charge as that of the tau in this cases. Hence 
combining the four momenta of this track with that of $W$ one can obtain a $W^{\pm}\stau^{\mp}_1$ invariant mass 
distribution peaking at the $\snu_{\tau_L}$ mass ($m_{\snu_{\tau_L}}$).This requires identification of the charge of 
a jet out of a tau decay, the efficiency of which has been assumed to be 100\%.   The events in which the tau has
originated from the decay of a chargino ($\chi^{\pm}_1 \r \snu_{\tau_L}\tau^{\pm}$), and the tau out of a neutralino 
decay goes unidentified, will contribute to the background events in this method, which we will discuss in the next 
subsection. Nevertheless, this method works very well in determining the correct $W^{\pm}\stau^{\mp}_1$ pair and one 
obtains the invariant mass peak at the correct value of the corresponding mass of the left-chiral sneutrino 
($m_{\snu_{\tau_L}}$). Furthermore, this method can be used irrespective of the possibility of reconstruction of the 
neutralino and chargino mass\footnote{It should be mentioned here, that the mass associated with the charge track, which 
can be found following our earlier work \cite{Biswas:2009zp}, is an inevitable input for both the methods.}.

\item \underline{Using chargino-neutralino mass information (CNMI):} Though the method based on OSCT does not depend on 
the 
reconstructability of other superparticle masses, its main disadvantage is that it not only reduces half of the signal 
event but also includes many background events. The correct $W^{\pm}\stau^{\mp}_1$ pair can also be determined if one 
uses the information of the chargino or/and neutralino mass, by 
looking at the end point in the invariant mass distribution of the track and tau-jet pair. The right combination will have
the end point at the corresponding neutralino mass ($m_{\chi^0_i},~i=1,2$). In cases where the tau out of a neutralino 
decay goes undetected and the tau out of a chargino decay gets identified, then the $\stau-W-\tau_{j}$ pair invariant mass 
distribution will have a end point at the corresponding chargino mass ($m_{\chi^{\pm}_1}$). We have combined the $W^{\pm}$ 
with the corresponding track ($\stau^{\mp}_1$) when either of the $m_{\chi^0_i}-M_{\stau\tau_j}\leq20~GeV$ or 
$m_{\chi^{\pm}_i}-M_{\stau W \tau_j}\leq20~GeV$ criterion is satisfied.

\end{enumerate}


\subsection{ Backgrounds and event selection criteria}

In this subsection we discuss the possible SM backgrounds that can fake our signal, namely, 
$W+\tau_j+2\stau+E_{T}\miss+X$ and prescribed the requisite cuts to minimise them. First of all, 
we have advocated the following basic cuts for each event to validate our desired final state:

\begin{itemize}
\item $ p_{T}^{lep}, ~p_T^{track} > 10$~GeV
\item $ p_{T}^{hardest-jet} > 75$~GeV 
\item $ p_{T}^{other-jets} > 30$~GeV 
\item $ \sla{E_T}> 40 $~GeV
\item $|\eta| < 2.5$  for leptons, jets and stau 
\item $\Delta R_{ll}>0.2,~ \Delta R_{lj}>0.4$, where $\Delta R^2=\Delta \eta^2+\Delta \phi^2$
\item $\Delta R_{\stau l}>0.2, ~\Delta R_{\stau j}>0.4$
\item $ \Delta R_{jj} >0.7$
\end{itemize}

\begin{figure}[ht]
\centerline{\epsfig{file=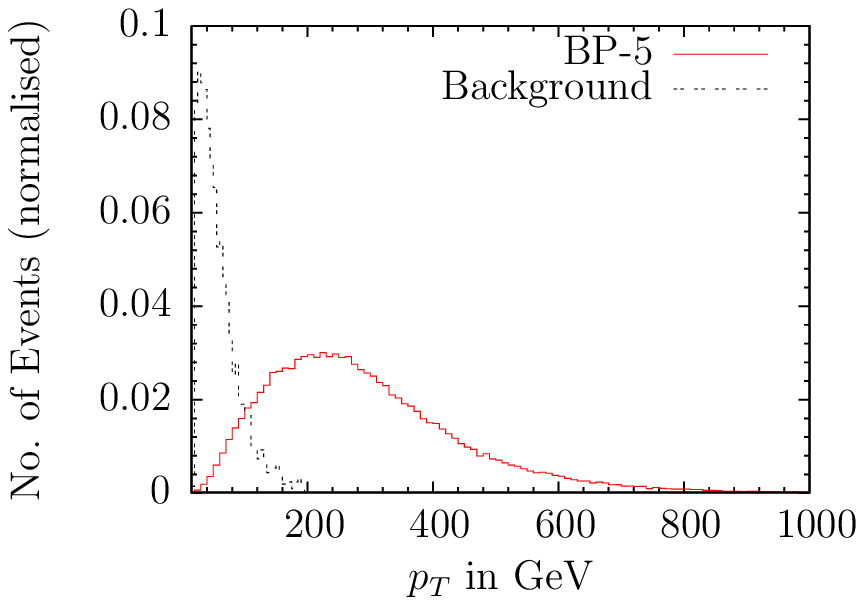,width=7.0cm,height=6.0cm,angle=-0}
\hskip 20pt \epsfig{file=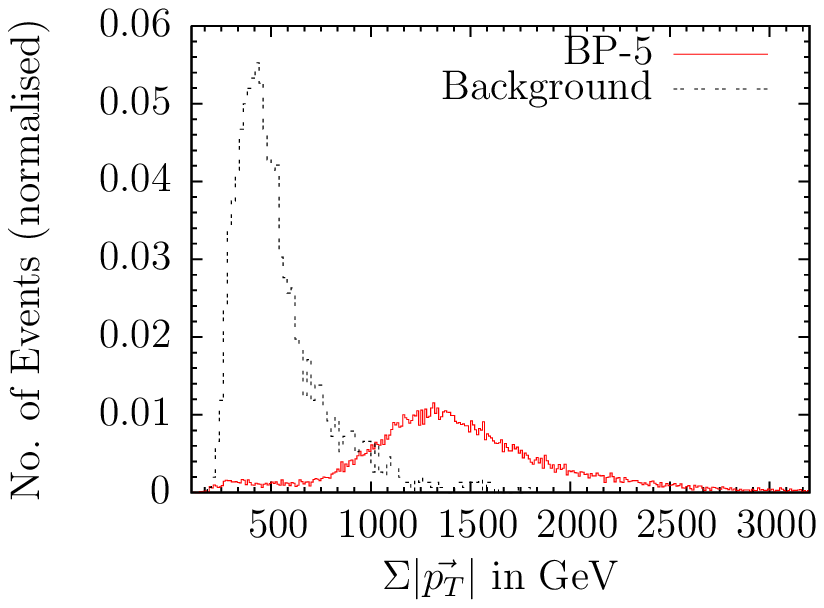,width=7.0cm,height=6.0cm,angle=-0}}
\vskip 10pt
\caption{\small \it {$p_T$ of the harder muonlike track (left) and  $\Sigma |\vec{p_T}|$ (right)
distribution (normalised to unity) for the signal (BP5) and the background, 
with $E_{cm}$=10 TeV.}} 
\end{figure}

\begin{figure}[h]
\centerline{\epsfig{file=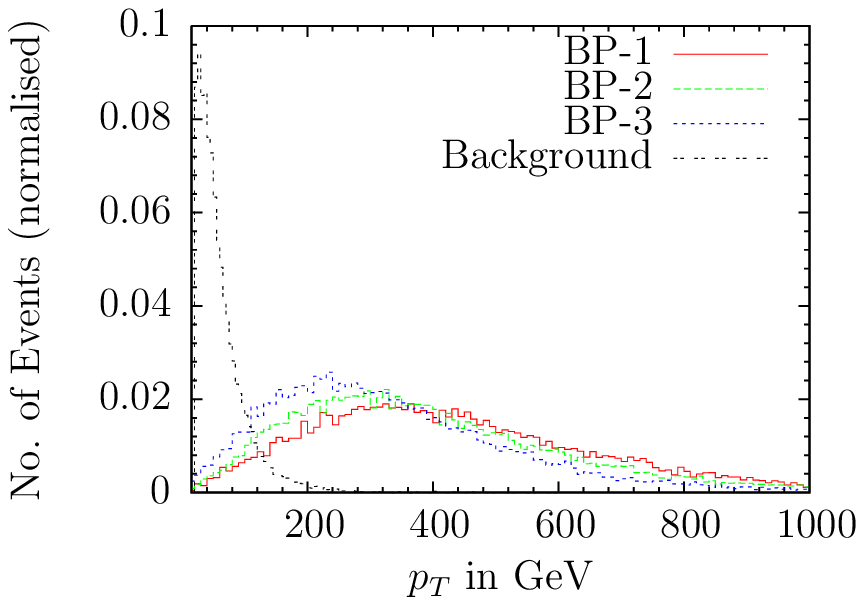,width=7.0cm,height=6.0cm,angle=-0}
\hskip 20pt \epsfig{file=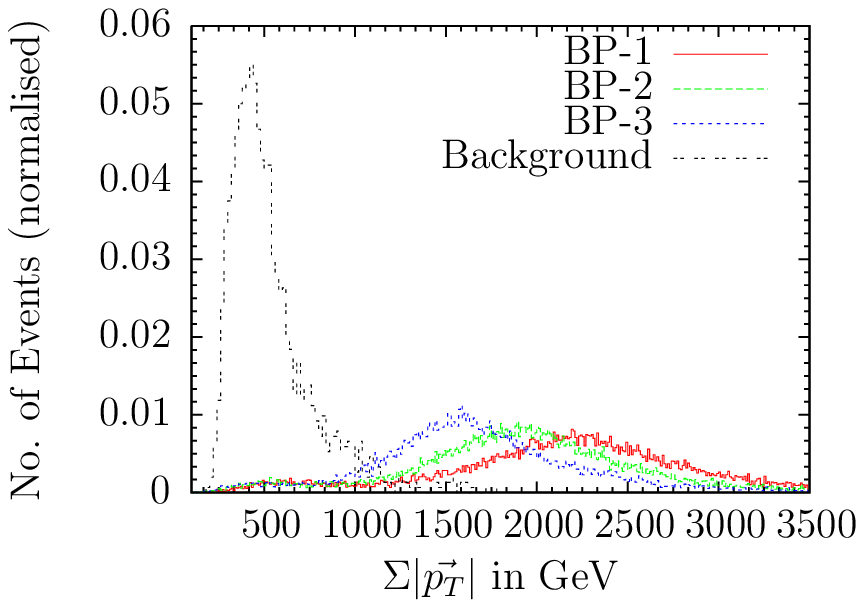,width=7.0cm,height=6.0cm,angle=-0}}
\vskip 10pt
\caption{\small \it {$p_T$ of the harder muonlike track (left) and  $\Sigma |\vec{p_T}|$ (right)
distribution (normalised to unity) for the signal (BP1, BP2 and BP3) and 
the background, with $E_{cm}$=14 TeV.}}
\end{figure}

\vspace{-0.25cm}
\begin{figure}[hp]
\centerline{\epsfig{file=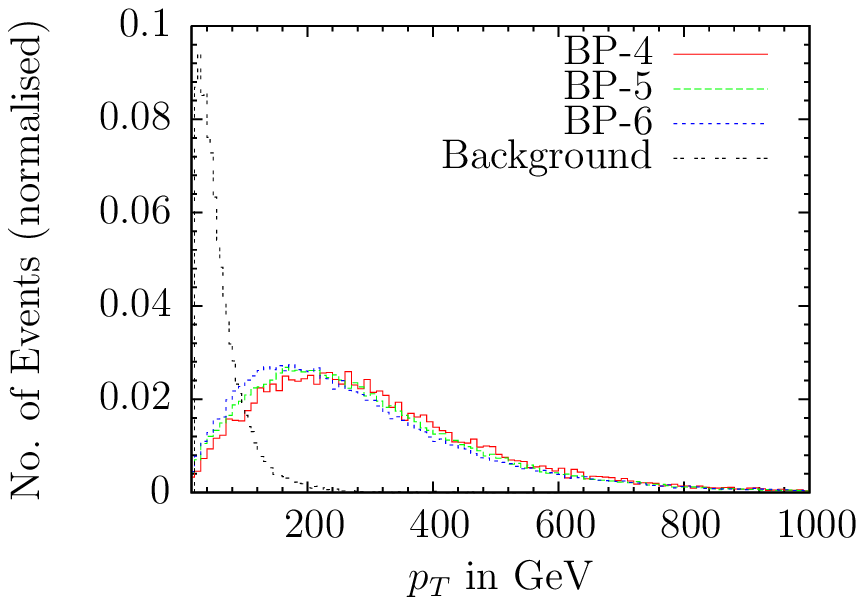,width=7.0cm,height=6.0cm,angle=-0}
\hskip 20pt \epsfig{file=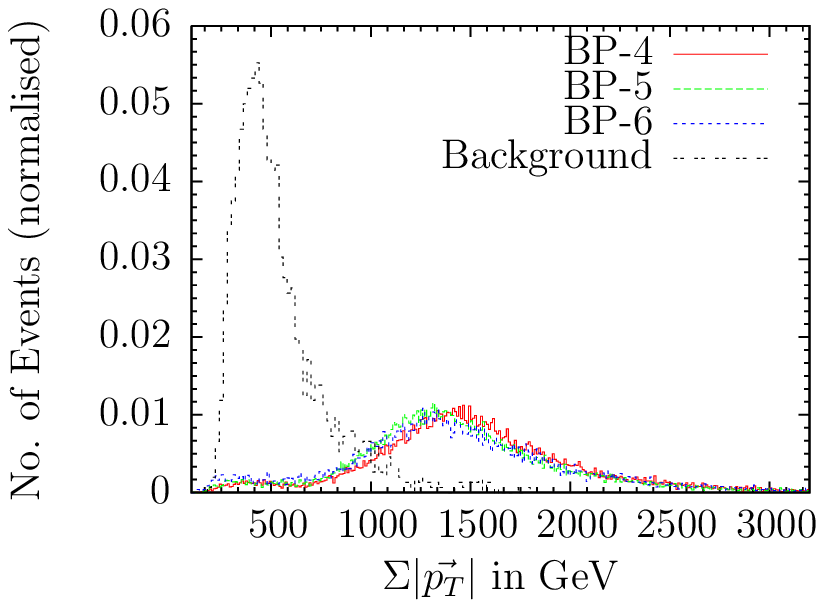,width=7.0cm,height=6.0cm,angle=-0}}
\vskip 10pt
\caption{\small \it {$p_T$ of the harder muonlike track (left )and  $\Sigma |\vec{p_T}|$ (right)
distribution (normalised to unity) for the signal (BP4, BP5 and BP6) and 
the background, with $E_{cm}$=14 TeV.}}
\end{figure}
\vspace{0.5cm}

 The above cuts are applied for simulation with both the center-of-mass energies of 10 and 14 TeV. 
Keeping detector resolutions of the momenta of each particle in mind, different Gaussian resolution functions
have been used for electrons, muons/staus and jets. These exactly follow the path prescribed in our earlier work 
\cite{Biswas:2009zp}. We have assumed a tau identification efficiency of 50\% for one prong decay of a tau lepton, 
whereas a rejection factor of 100 has been used for the non-taujets \cite{Rainwater:1998kj,Asai:2004ws, CMS1}. 
To identify the $W$ in its  hadronic decay mode, we have used the following criteria:

\begin{itemize}
\item the invariant mass of any two jets should lie within $M_W-20<M_{jj}<M_W+20$
\item the separation between the stau and the direction formed out of the vector sum of the
momenta of the two jets (produced in $W$ decay) should lie within $\Delta R=0.8$
\end{itemize}

The charged track of a massive particle in the muon chamber can be faked by the muon itself.
In our earlier study, we have shown that the SM backgrounds like $t\bar t$, $ZZ$, $ZW$, $ZH$ can contribute to 
the $\tau_j+2\stau+E_{T}\miss+X$ final state. In the present study we have an additional $W$ in the final state. 
This requirement further reduces the contribution from SM processes. Above all, 
demanding an added degree of hardness of the charged tracks and a minimum value of the 
scalar sum of transverse momenta ($\Sigma |\vec{p_T}|$) of all the visible particle in the final state, the 
contribution from above SM processes to the $W+\tau_j+2\stau+E_{T}\miss+X$ final state can be suppressed
considerably. For the simulation at $E_{cm}=10 ~TeV$, we have adopted the following cuts on 
the $p_T^{\stau}$ and $\Sigma |\vec{p_T}|$ variables (See Figure 1):

\begin{itemize}
\item $ p_T^{track} > 75$~GeV
\item $\Sigma |\vec{p_T}| > 700$~GeV
\end{itemize}

\noindent{
where as, for the simulation at a higher center of mass energy $E_{cm}=14 ~TeV$ the degree of hardness raised 
to(See Figure 2 and 3):}

\begin{itemize}
\item $ p_T^{track} > 100$~GeV
\item $\Sigma |\vec{p_T}| > 1$~TeV
\end{itemize}

\section{ Results and discussions}

We present the numerical results of our study in this section. In Table 3 we have presented the results 
of simulation at $E_{cm}=10 ~TeV$ for BP5, to illustrate that it is possible 
to reconstruct the left-chiral tau-sneutrino even at the early phase of the LHC run at an integrated luminosity 
of $3~fb^{-1}$ (Figure 4). The results of simulation at $E_{cm}=14 ~TeV$ for two different luminosities have also been shown 
in Table 4 and 5. Depending on the luminosity and available center of mass energy, it is possible to probe some or 
all of the benchmark points we have studied, at the LHC.

The numerical results enable us to assess the relative merits of the OSCT and CNMI methods.
It is obvious from Table 3-5 that we have larger number of events when the chargino-neutrlino mass information 
(CNMI) rather than the opposite sign charged tracks (OSCT) has been used to reconstruct the mass 
of the $\snu_{\tau_L}$, as one would expect the number of events to be less if one is restricted
to opposite sign charged tracks. Also, the chance 
of including the wrong $W^{\pm}\stau^{\mp}_1$-pair is more in the OSCT method.
For example, as has already been mentioned, $\chi^{\pm}_1\chi^{0}_1$ produced in cascades
can give rise to $W+\tau_j+2\stau+E_{T}\miss+X$ final state, where the tau out of a chargino decay 
($\chi^{\pm}_1 \r \snu_{\tau_L}\tau^{\pm}$) has been identified and not the one originated from a 
neutralino decay. In this situation one ends up with a wrong combination of $W^{\pm}\stau^{\mp}_1$
pair if one is using OSCT. The contribution of such background, however, is not counted
when CNMI is used. In spite of this, the fact that no information on chargino and neutralino masses is used in OSCT,
is of advantage in independently confirming the nature of the spectrum. Both the methods discussed above are prone to
background contamination within the model
itself, such as misidentification of $W$ and the decay like $\chi^{\pm}_{1/2}\r\chi^{0}_{1/2}W^{\pm}\r
\stau_1W^{\pm}\tau$, which  smear the peak, as is visible from Figure 5 and 6.

From the numerical results, it is seen that the number of events start getting increased as one moves from BP1 
to BP5, due to the fact that the production cross section of $\snu_{\tau_L}$ in cascade decay of squarks 
and/or gluinos get enhanced. However, at BP6 one has less number of events. The reason is twofold: First,
the decay branching fraction of $\snu_{\tau_L}\r\stau_1W$
reduces from 84\% for BP1 to 60\% for BP5. The enhanced production cross-section thus has dominated effect.
However, for BP6 the branching ratio falls to 34.5\%, thus affecting the event rates adversely. Secondly, 
the mass difference $m_{\snu_{\tau_L}}-(m_{\stau_1}+m_W)$ is of the order of 20 GeV, which restricts the stau 
track from passing the requisite hardness cut for a sizeable number of events. Also, the $\stau_1W$ invariant 
mass peak is badly affected at BP6, as the contribution from the decay $\chi^{\pm}_{1/2}\r\chi^{0}_{1/2}W^{\pm}$ 
is maximum due to the increase in the branching ratio. The number of events within a bin of $\pm 20~GeV$ around 
the peak obtained using CNMI is comparable to that obtained using OSCT at this benchmark point. This is due to
the fact that in case of mass reconstruction using CNMI, the information about the lightest neutralino mass is
not available at BP6 \cite{Biswas:2009zp}.

We have also explored the mSUGRA parameter space to study the feasibility of sneutrino reconstruction. 
A thorough scan over the $m_0-M_{1/2}$ plane
has been done using the  spectrum generator {\bf \begin{footnotesize} SuSpect v2.34\end{footnotesize}} \cite{SUSPECT}, 
which leads to a $\stau$ LSP in 
a usual mSUGRA scenario without the right handed sneutrino and identified the region where it is possible to
reconstruct the left-chiral stau neutrino with more than 25 events within the vicinity of the $\snu_{\tau_L}$ 
mass peak at an integrated luminosity of $30~fb^{-1}$. The corresponding plots are depicted in Figure 7.

The regions where reconstruction is possible have been determined using
the following criteria:

\newpage

\begin{table}[htbp]
\begin{tabular}{||c|c|c|c|c|c|c||}
\hline
\hline
  & \multicolumn{2}{c|} {basic cuts} & \multicolumn{2}{c|} {$p_T$+$\Sigma{|p_T|}$} & 
     \multicolumn{2}{c||} {$|M_{peak}-M_{\stau W}|\le 20$} \\
\hline
 {\bf BP5} & CNMI & OSCT & CNMI & OSCT & CNMI & OSCT \\
  & 132 & 84 & 107 & 66 & 23 & 15 \\
\hline 
\hline
\end{tabular}
\caption{\small \it {Number of signal events for the 
$W+\tau_j+2\stau$ (charged-track)+$E_{T}\miss+X$ final state, 
considering all SUSY processes for BP5 with $E_{cm}$=10 TeV at an 
integrated luminosity of 3 $fb^{-1}$ assuming tau identification 
efficiency $\epsilon_{\tau}=50\%$. In Table CNMI stands for reconstruction 
of $m_{\snu_{\tau_L}}$ using Chargino-Neutralino Mass Information whereas OSCT 
implies reconstruction of the same using Opposite Sign Charged Tracks.}}

\label{tab:4}
\end{table}
\begin{figure}[htbp]
\centerline{\epsfig{file=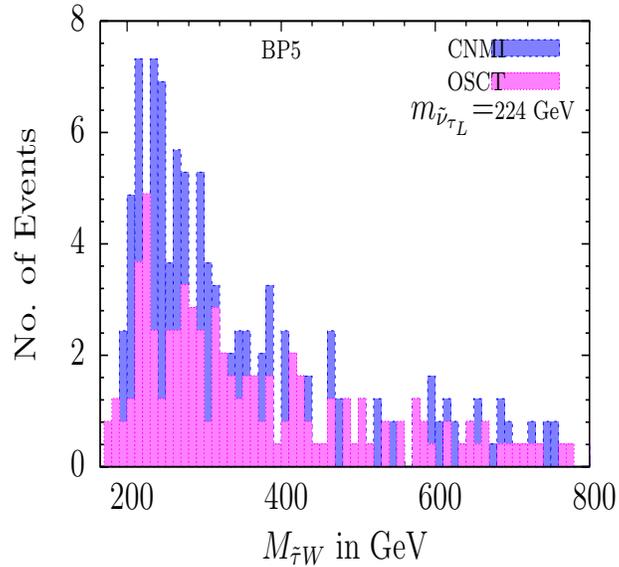,width=8.0cm,height=7.5cm,angle=-0}}
\caption{\small \it {The invariant mass ($M_{\stau W}$) distribution in 
$W+\tau_j+2\stau$ (charged-track)+$E_{T}\miss+X$ final state for BP5 
assuming tau identification efficiency $(\epsilon_{\tau})=50\%$ at an 
integrated luminosity of 3 $fb^{-1}$ and with $E_{cm}$=10 TeV. In Figure 
CNMI stands for reconstruction of $m_{\snu_{\tau_L}}$ using Chargino-
Neutralino Mass Information whereas OSCT implies reconstruction of 
the same using Opposite Sign Charged Tracks.}} 
\end{figure}

\renewcommand{\baselinestretch}{1.1} 
\begin{table}[htbp]
\begin{tabular}{||l|c|c|c|c|c|c||}
\hline
\hline
  &\multicolumn{2}{c}{\bf BP1}&\multicolumn{2}{c}{\bf BP2}&\multicolumn{2}{c||}{\bf BP3}\\
  &CNMI &OSCT &CNMI &OSCT &CNMI &OSCT \\
\hline
basic cuts & 318 & 248 & 984 & 715 & 4060 & 2550 \\
\hline
$p_T$+$\Sigma{|p_T|}$ cut& 250 & 205 & 775 & 574 & 3030 & 1904 \\
\hline
$|M_{peak}-M_{\stau W}|\le 20$& 62 & 38 & 187 & 119 & 634 & 364 \\
\hline 
\hline
\end{tabular}
\caption{\small \it {Number of signal events for the $W+\tau_j+2\stau$ 
(charged-track)+$E_{T}\miss+X$ final state, considering all SUSY processes, 
for BP1, BP2 and BP3  with $E_{cm}$=14 TeV at an integrated luminosity of 100 
$fb^{-1}$ assuming tau identification efficiency $\epsilon_{\tau}=50\%$. In 
Table CNMI and OSCT stand for the same as in Table 3.}}
\label{tab:5}
\end{table}
\renewcommand{\baselinestretch}{1.1} 
\begin{table}[htbp]
\begin{tabular}{||l|c|c|c|c|c|c||}
\hline
\hline
  &\multicolumn{2}{c}{\bf BP4}&\multicolumn{2}{c}{\bf BP5}&\multicolumn{2}{c||}{\bf BP6}\\
  &CNMI &OSCT &CNMI &OSCT &CNMI &OSCT \\
\hline
basic cuts & 8634 & 5450 & 12519 & 8014 & 4013 & 3995 \\
\hline
$p_T$+$\Sigma{|p_T|}$ cut& 6145 & 3709 & 8654 & 5363 & 2560 & 2537 \\
\hline
$|M_{peak}-M_{\stau W}|\le 20$& 1218 & 689 & 1471 & 866 & 385 & 349 \\
\hline 
\hline
\end{tabular}
\caption{\small \it {Number of signal events for the $W+\tau_j+2\stau$ 
(charged-track)+$E_{T}\miss+X$ final state, considering all SUSY processes, 
for BP4, BP5 and BP6  with $E_{cm}$=14 TeV at an integrated luminosity of 100 
$fb^{-1}$ assuming tau identification efficiency $\epsilon_{\tau}=50\%$. In 
Table CNMI and OSCT stand for the same as in Table 3.}}
\label{tab:6}
\end{table}
\begin{figure}[htbp]
\centerline{\epsfig{file=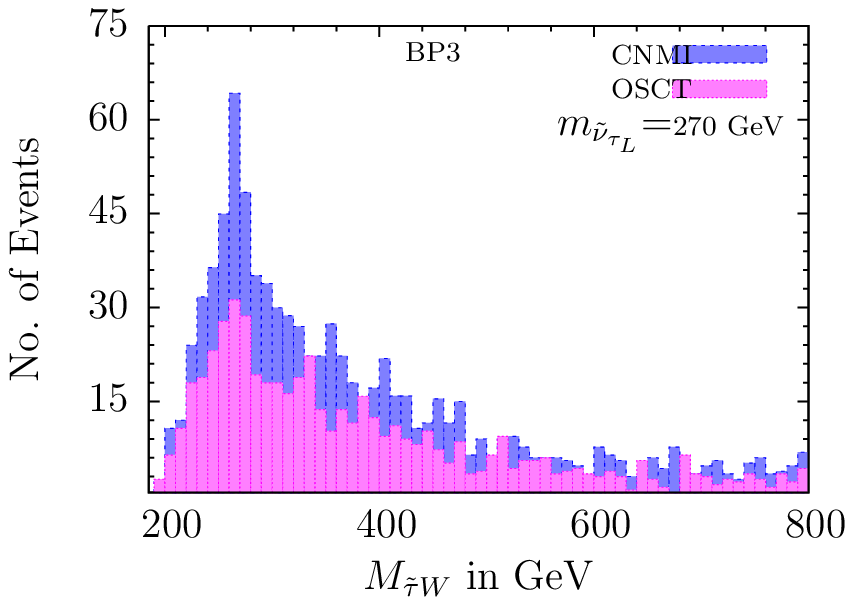,width=7.0cm,height=6.0cm,angle=-0}
\hskip 20pt \epsfig{file=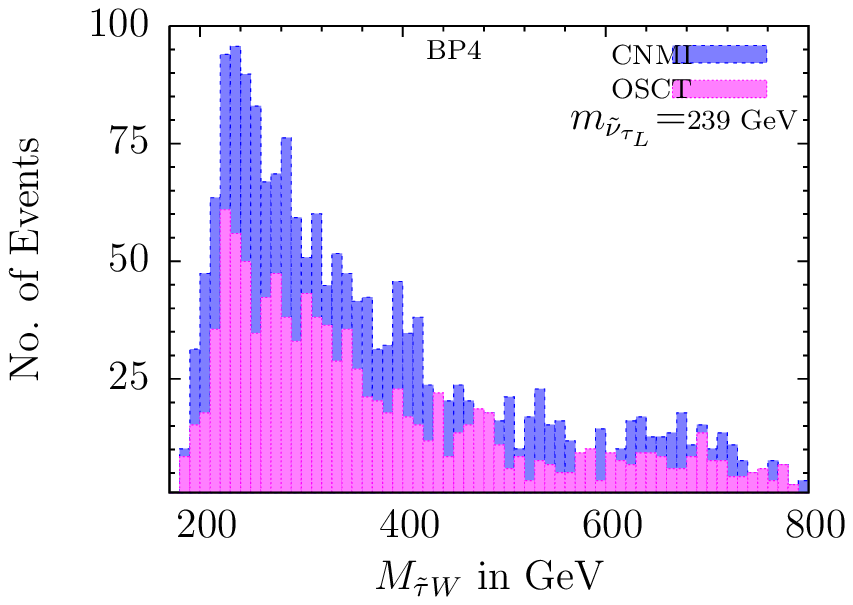,width=7.0cm,height=6.0cm,angle=-0}}
\vskip 10pt
\centerline{\epsfig{file=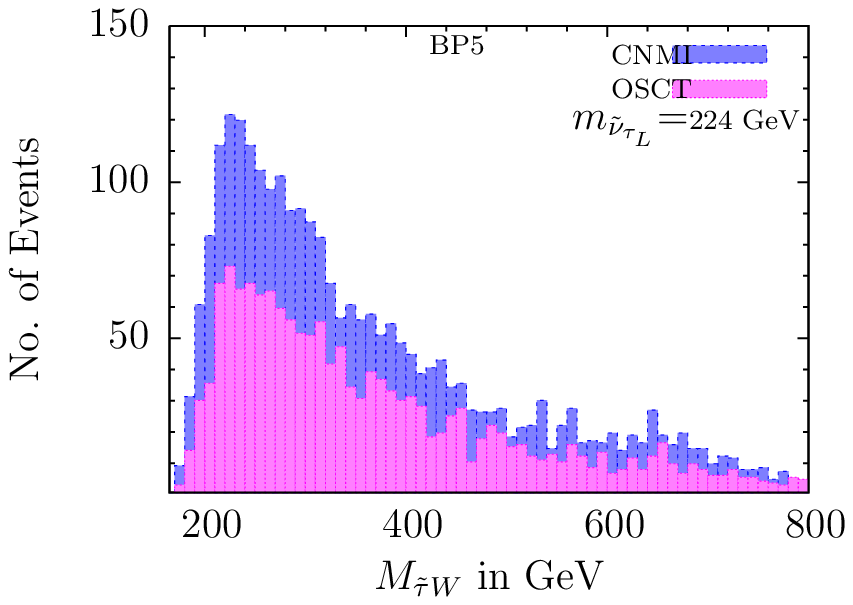,width=7.0cm,height=6.0cm,angle=-0}
\hskip 20pt \epsfig{file=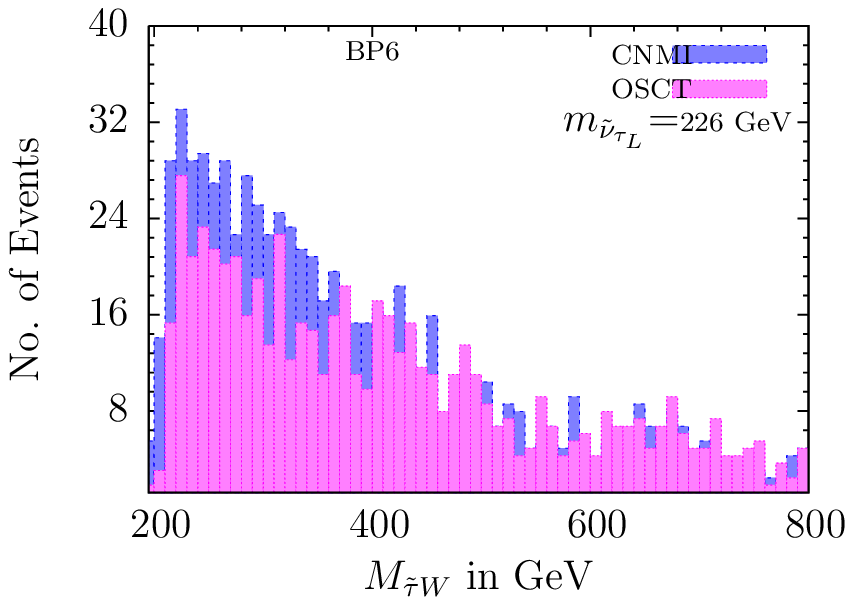,width=7.0cm,height=6.0cm,angle=-0}}
\vskip 20pt
\caption{\small \it {The invariant mass ($M_{\stau W}$) distribution in 
$W+\tau_j+2\stau$ (charged-track)+$E_{T}\miss+X$ final state for four of 
our proposed benchmark points assuming tau identification efficiency 
$(\epsilon_{\tau})=50\%$ at an integrated luminosity of 30 $fb^{-1}$ and 
centre of mass energy 14 $TeV$. In Figure CNMI and OSCT stand for the 
same as in Figure 4.}} 
\end{figure}


\begin{figure}[htbp]
\centerline{\epsfig{file=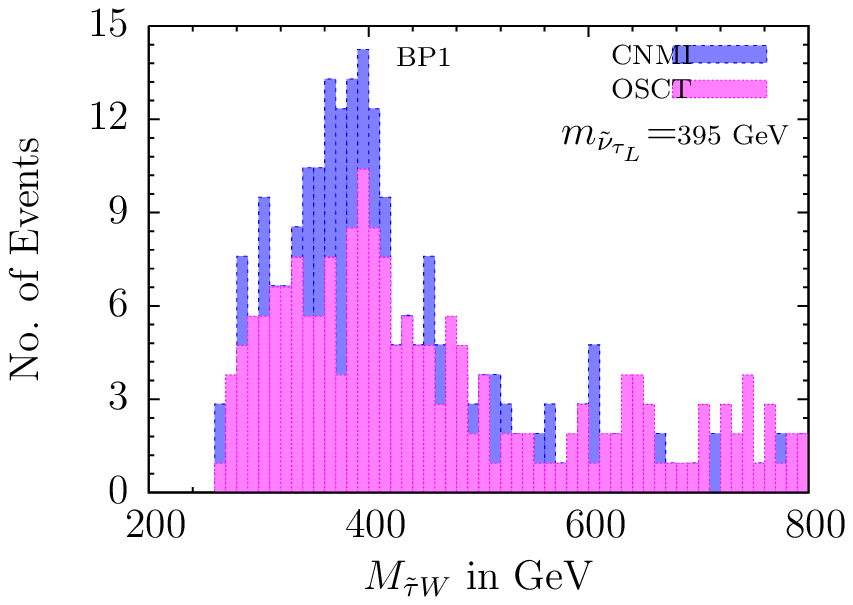,width=7.0cm,height=6.0cm,angle=-0}
\hskip 20pt \epsfig{file=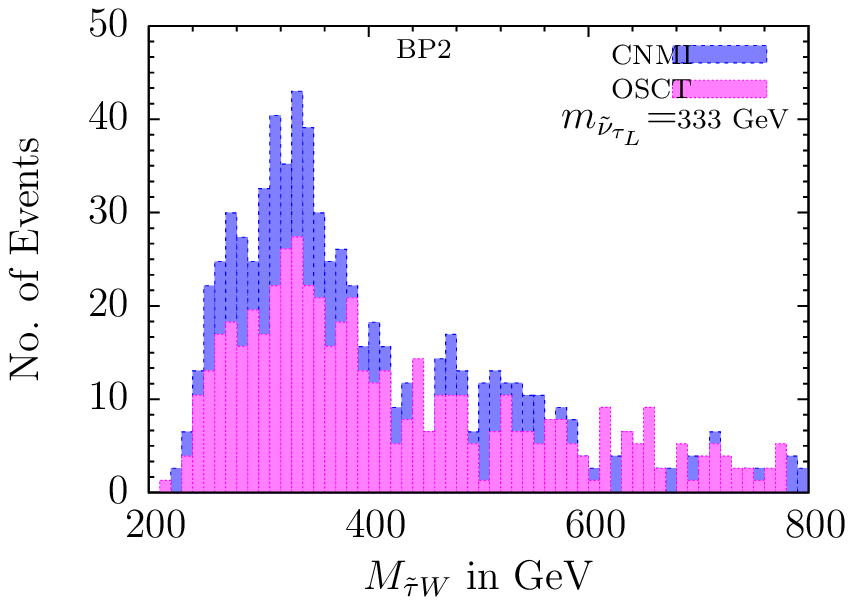,width=7.0cm,height=6.0cm,angle=-0}}
\vskip 10pt
\centerline{\epsfig{file=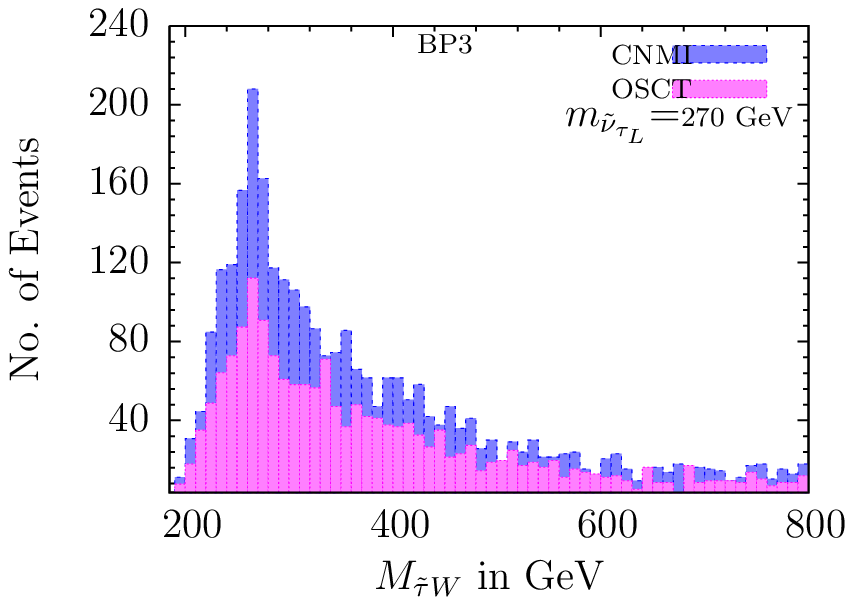,width=7.0cm,height=6.0cm,angle=-0}
\hskip 20pt \epsfig{file=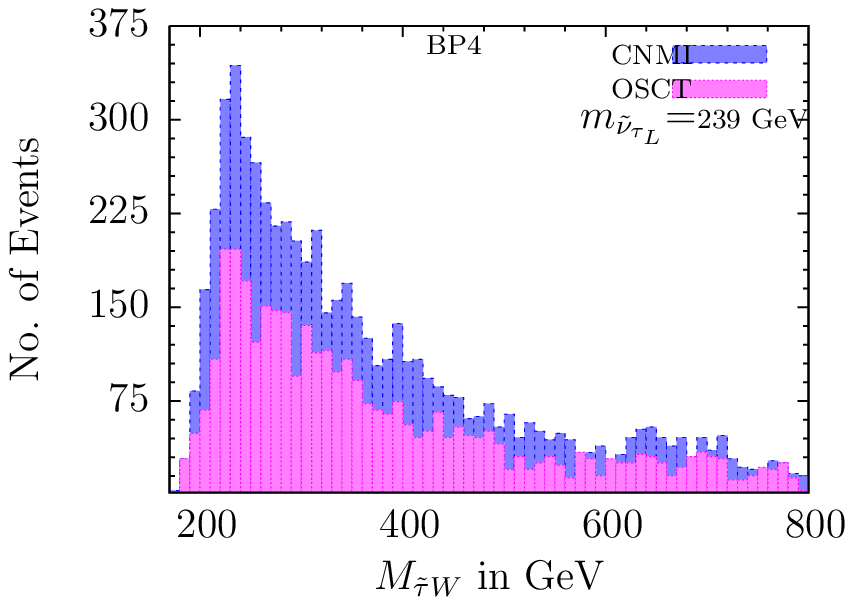,width=7.0cm,height=6.0cm,angle=-0}}
\vskip 10pt
\centerline{\epsfig{file=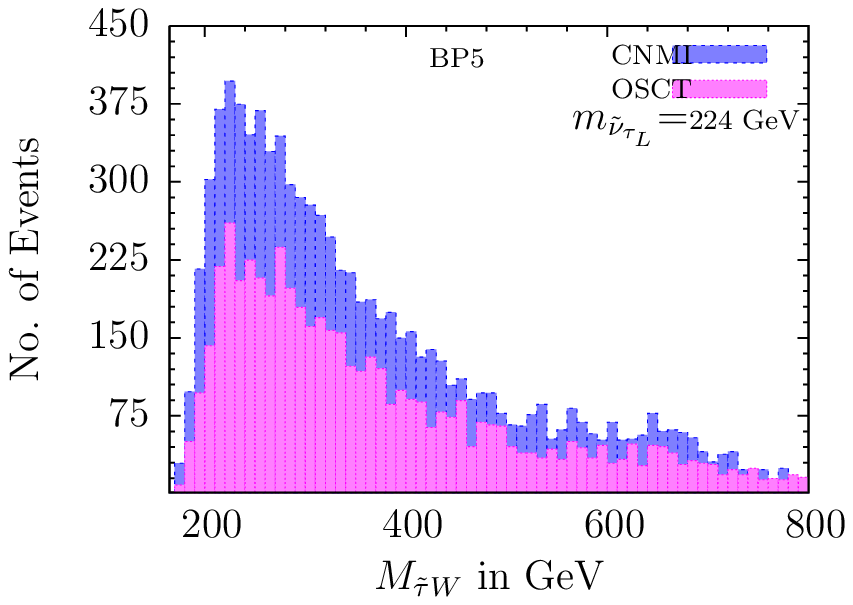,width=7.0cm,height=6.0cm,angle=-0}
\hskip 20pt \epsfig{file=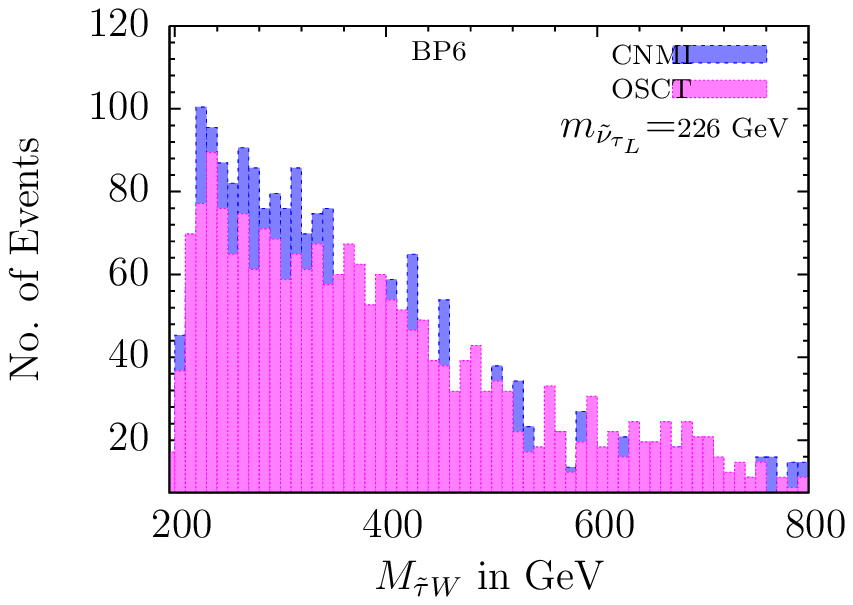,width=7.0cm,height=6.0cm,angle=-0}}
\vskip 20pt
\caption{\small \it {The invariant mass ($M_{\stau W}$) distribution in 
$W+\tau_j+2\stau$ (charged-track)+$E_{T}\miss+X$ final state
for all the benchmark points assuming tau identification efficiency 
$(\epsilon_{\tau})=50\%$ at an integrated luminosity of 100 $fb^{-1}$ and 
centre of mass energy 14 $TeV$. In Figure CNMI and OSCT stand for the 
same as in Figure 4.}} 
\end{figure}


\begin{itemize}

\item In the parameter space, we have not gone into regions where the gluino 
mass exceeds $\approx 2 ~TeV$.

\item The number of events within a bin of $\pm20~GeV$ around 
the peak must be greater than a specific number for the corresponding luminosity.

\item Also it is possible to reconstruct at least one of the neutralinos at the region 
of $m_0-M_{1/2}$ under consideration, when the mass information is used, as a criterion 
of finding correct $W^{\pm}\stau^{\mp}_1$. We should mention here that, for this study
in identifying the region of mSUGRA parameter space, we have used  neutralino mass 
information alone and not the information on chargino mass.

\end{itemize}

\begin{figure}[htp]
\begin{center}
\centerline{\epsfig{file=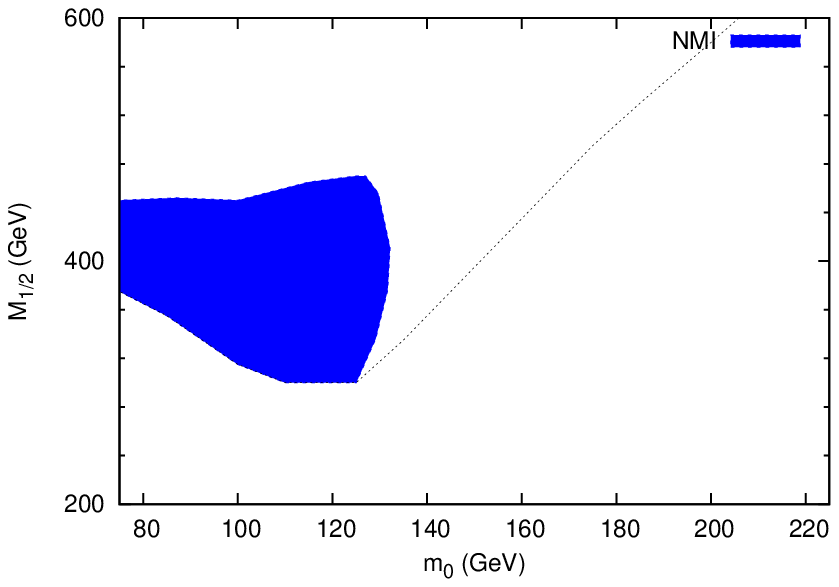,width=7.5cm,height=6.0cm,angle=-0}
\hskip 20pt \epsfig{file=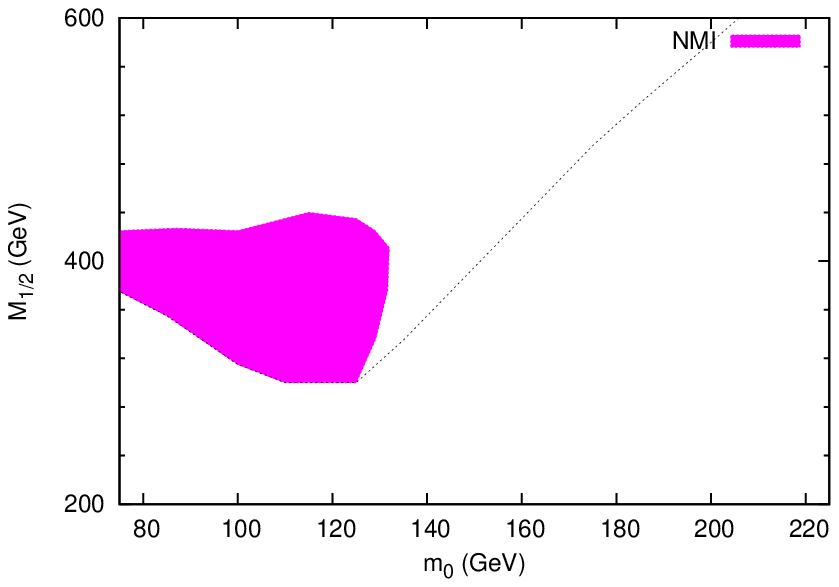,width=7.5cm,height=6.0cm,angle=-0}}
\caption{The region  in the $m_0-M_{1/2}$ plane (with $tan\beta=30$ and $A_0=100$), where it is possible 
to reconstruct the left-chiral sneutrino at an integrated luminosity of 30 $fb^{-1}$ and center-of-mass 
energy 14 $TeV$ with more than 25 events in the vicinity of the peak. In the Figure blue (dark shade) 
region represents reconstruction of $m_{\snu_{\tau_L}}$ using only Neutralino Mass Information (NMI), while
the pink region (light shade) stands for reconstruction of the same using Opposite Sign Charged Tracks (OSCT)
for the $W+\tau_j+2\stau$ (charged-track)+$E_{T}\miss+X$ final state. The entire region above the dashed line 
indicates the scenario where one has a $\td{\nu_R}$-LSP and a $\stau$-NLSP.}
\end{center}
\end{figure}
\vspace{-0.5cm}

\section{Summary and conclusions}

We have considered a SUSY scenario where the LSP is dominated by a right-sneutrino
state, while a dominantly right-chiral stau is the NLSP. The stau, being
stable on the length scale of collider detectors, gives rise to
charged track of massive particle at the muon chamber. It is also
shown that such a scenario follows naturally from a high-scale scenario
of universal scalar and gaugino masses. 

We have investigated the possibility of reconstruction of the left-chiral
tau sneutrino in such a scenario. For that we have studied the final state 
consisting of $W+\tau_j+2\stau+E_{T}\miss+X$.
We have also prescribed two different strategies for the reconstruction of the 
mass of the $\snu_{\tau_L}$. One is independent of the reconstructability of other particles
as it does not uses the mass information and the other one does depend on the reconstructability
of the chargino and neutralino masses. The cuts imposed on the kinematic variables to eliminate the 
SM backgrounds are motivated by our recent studies on chargino and neutralino reconstruction 
under similar circumstances. We have demonstrated the feasibility of reconstructing the mass of the $\snu_{\tau_L}$
even at the early phase of LHC run with $E_{cm}=10 ~TeV$ and at an integrated luminosity of $3~fb^{-1}$ for a  
particular benchmark point (BP5) for illustration. The results for simulation with higher LHC center-of-mass energy 
$E_{cm}=14 ~TeV$ at two different luminosities have also been shown.

A thorough scan over the $m_0-M_{1/2}$ plane has been performed in this study, 
which shows that a significant region of the mSUGRA parameter space can be probed at the
LHC with sufficient number of events at an integrated luminosity of $30~fb^{-1}$ with 
$E_{cm}=14 ~TeV$ using our prescribed methods.

To conclude, the MSSM with a right-chiral sneutrino superfield for each generation is a worthful possibility to 
look at the LHC. It not only offers a distinct SUSY signal in the form of a pair of charged tracks of massive
particles but also opens a new vista in the reconstruction of the superparticle masses.

{\bf Acknowledgement:}  The author thanks Biswarup Mukhopadhyaya for helping in the 
preparation of this manuscript and giving valuable suggestions. This work is partially supported 
by funding available from the Department of Atomic Energy, Government of India for the Regional 
Center for Accelerator-based Particle Physics, Harish-Chandra Research Institute. Computational 
work for this study was partially carried out at the cluster computing facility of Harish-Chandra 
Research Institute ({\tt http:/$\!$/cluster.mri.ernet.in}).


\end{document}